\documentclass[11pt,twoside]{book} 
\usepackage{asp2008s}
\usepackage{times}
\usepackage{lscape}
\usepackage{epsf}

\usepackage{graphicx}
\usepackage{amssymb}
\usepackage{longtable}
\usepackage[figuresright]{rotating}
\usepackage{multirow}

\newcommand{\simless}{\mathbin{\lower 3pt\hbox {$\rlap{\raise 5pt\hbox{$\char'074$}}\mathchar"7218$}}}

\mainmatter

\newlength{\deftabcolsep}
\begin{document}

\title{The Multiwavelength Picture of Star Formation in the Very\\ Young Open Cluster NGC\,6383}
\author{Gregor Rauw and Micha\"el De Becker}
\affil{Institut d'Astrophysique et de G\'eophysique, Universit\'e de Li\`ege,\\ All\'ee du 6 Ao\^ut 17, B\^at B5c, B-4000 Li\`ege, Belgium}


\begin{abstract}
We review the properties of the very young ($\sim 2$\,Myr) open cluster NGC\,6383. The cluster is dominated by the massive binary HD\,159176 (O7\,V + O7\,V). The distance to NGC\,6383 is consistently found to be $1.3 \pm 0.1$\,kpc and the average reddening is determined to be $E(\bv) = 0.32 \pm 0.02$. Several pre-main sequence candidates have been identified using different criteria relying on the detection of emission lines, infrared excesses, photometric variability and X-ray emission.
\end{abstract}


\section{NGC\,6383 and its Surroundings}
NGC\,6383 ($\alpha_{2000} = 17^h34^m48^s$, $\delta_{2000} = -32^{\circ}34\farcm0$; $l_{II} = 355.69^{\circ}$, $b_{II} = +0.04^{\circ}$) is a rather compact open cluster which could be part of the Sgr\,OB1 association together with NGC\,6530 and NGC\,6531. The cluster was originally discovered by John Herschel in 1834, and listed as h\,3689 in Herschel's Cape Catalogue published in 1847. Most probably due to a clerical error, it also appears in Dreyer's New General Catalogue as NGC\,6374, an identifier that is however not commonly used.

\begin{figure}[htb]
\centering
\includegraphics[draft=False,width=\textwidth]{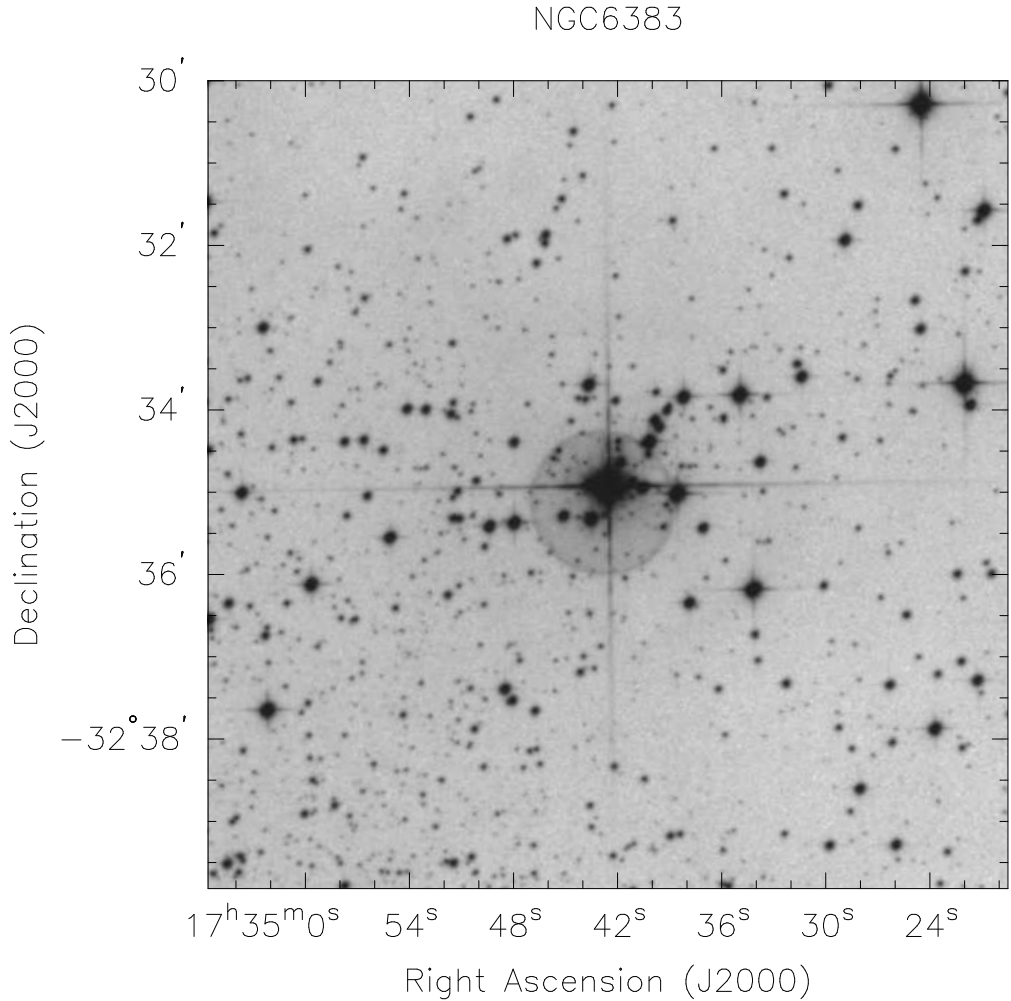}
\caption{The region around the core of NGC\,6383 as seen on the UKSTU survey red plate. The bright star in the middle of the field of view is the massive binary system HD\,159176. The field size is 10\,arcmin $\times$ 10\,arcmin.\label{optical}}
\end{figure}

Trumpler \cite{Trumpler} pointed out that NGC\,6383 belongs to a category of clusters in which a few dozen faint stars are closely grouped around a bright hot central star that frequently happens to be a binary system. In the case of NGC\,6383, the cluster is centered on the O-type binary HD\,159176 (m$_V$ = 5.7, see Fig.\,\ref{optical}) that dominates the emission from the cluster over a broad range of energies from the near-infrared to the X-ray domain. Apart from HD\,159176, the cluster currently harbours no stars earlier than B1, although de Wit et al.\ \cite{deWit} suggested that the eclipsing binary candidate HD\,158186 (O9.5\,V, Marchenko et al.\ 1998) might have been ejected from NGC\,6383 through dynamical interactions in the cluster core.

NGC\,6383 and more specifically HD\,159176 are likely responsible for the ionization of the H\,{\sc ii} region RCW\,132 (Rodgers, Campbell \& Whiteoak 1960) also known as S\,11 (Sharpless 1953)\footnote{Note that the S\,11 identifier is not to be confused with the number of the nebula in the revised catalogue of H\,{\sc ii} regions published by Sharpless \cite{Sharp2}. In the latter catalogue, the nebula corresponds to entry number 12 (in the SIMBAD database, this identifier is referred to as Sh 2-012).} or Stromlo\,67 (Gum 1955). Rodgers et al.\ \cite{RCW} described RCW\,132 as a 110\,arcmin $\times$ 80\,arcmin medium brightness crescent-shaped region. Images of this emission nebula can be found for instance on plate 131 of Lyng\aa\ \& Hansson \cite{LH}.

During a survey of the galactic ridge at 1390\,MHz, Westerhout \cite{Westerhout} detected a large ($1.5^{\circ} \times 1.5^{\circ}$) ring-like radio structure roughly centered on NGC\,6383. However, Westerhout cautioned that this feature was difficult to separate from the background emission. Part of this source is likely due to RCW\,132.

\section{The Massive Binary HD\,159176 and Other Early-type Stars}
According to Lloyd Evans \cite{LE1}, NGC\,6383 harbours a total of 21 OB stars including HD\,159176 and V\,701\,Sco. The spectrum of HD\,159176 was described in the {\it Henry Draper Catalogue} as O5e with very wide lines that were almost double on some occasions. Trumpler \cite{Trumpler} was the first to report the double-lined spectroscopic binary (SB2) nature of the star and proposed orbital solutions for two possible periods of 4.920 and 3.368\,days. The first clear determination of the orbital period was performed by Conti, Cowley, \& Johnson \cite{conti} on the basis of optical data. The orbital solution was subsequently improved by Seggewiss \& de Groot \cite{segg}, Lloyd Evans \cite{LE2} and Stickland et al.\,\cite{stick}. Finally, the most recent orbital solution was published by Linder et al.\,\cite{Linder} based on a set of high-resolution optical echelle spectra analysed with a spectral disentangling method. The main result of these spectroscopic studies is that HD\,159176 consists of two very similar main-sequence stars of spectral type\footnote{The ((f)) tag indicates that the N\,{\sc iii} $\lambda\lambda$\,4634--40 lines are seen in emission whilst He\,{\sc ii} $\lambda$\,4686 is in absorption (see e.g.\ Walborn \& Fitzpatrick 1990).} O7((f))\,V, with the primary being slightly hotter than the secondary, revolving on a circular orbit with a period of 3.367\,d. Linder et al.\,\cite{Linder} also showed that the components of the system likely have their rotation synchronized with their orbital period.

The photometric monitoring of HD\,159176 reported by Thomas \& Pachoulakis \cite{TP} revealed low amplitude ($\sim$\,0.05 mag) ellipsoidal variations. However, the absence of photometric eclipses prevented these authors from deriving the inclination of the system. The thorough analysis by Pachoulakis\,\cite{pach} allowed him to determine stellar radii for both components of the order of 25\,\% of the orbital separation. The stars hence do not fill up their Roche lobe, contrary to the suggestion formulated by Conti et al.\ \cite{conti}.

Pfeiffer et al.\,\cite{pfeif} reported on the discovery of a UV emission component produced between the two stars, and interpreted this as a signature of a wind-wind interaction that is slightly wrapped around the secondary star. Pachoulakis\,\cite{pach} derived mass loss rates of 3.2\,$\times$\,10$^{-6}$ and 2.6\,$\times$\,10$^{-6}$\,M$_\odot$\,yr$^{-1}$ respectively for the primary and the secondary. These determinations were based on the fit of the C\,{\sc iv} $\lambda\lambda$ 1548-1551 and N\,{\sc v} $\lambda\lambda$ 1239-1243 lines with
synthetic profiles resulting from the combination of the individual profiles from both stars, and taking into account the effects of wind eclipses depending on the orbital phase. We note that, following a less sophisticated approach and using the same spectral lines, Howarth \& Prinja \cite{HP} derived a significantly lower mass loss rate of 6\,$\times$\,10$^{-7}$\,M$_\odot$\,yr$^{-1}$ for both stars.

HD\,159176 was detected as a rather bright X-ray emitter by the {\it EINSTEIN} (Chlebowski, Harnden, \& Sciortino 1989) and {\it ROSAT} (Bergh\"ofer, Schmitt, \& Cassinelli 1996) satellites. More recently, the observations performed with the {\it XMM-Newton} satellite shed new light on the X-ray emission from this close binary system (De Becker et al.\ 2004). HD\,159176 is a rather soft thermal X-ray emitter whose spectrum is dominated by plasma at temperatures of 2 -- 3 and 6 -- 10\,MK. Such temperatures are typical of plasma heated by shocks respectively within individual stellar winds and in colliding winds in close binaries. An additional indication of X-ray emission produced by the colliding winds comes from the X-ray luminosity that is too high to be explained by intrinsic X-ray emission from individual stellar winds alone. The modelling of the X-ray emission from the colliding winds revealed that the previous determinations of the mass loss rates (see above) may have been overestimated. Finally, De Becker et al.\,\cite{DeB} showed that purely thermal emission models failed to fit the X-ray spectrum up to 10.0 keV, although very good results were obtained using an additional power law to fit a high energy tail. This may point to a non-thermal emission component presumably produced by inverse Compton scattering of photospheric UV photons by relativistic electrons accelerated in the wind interaction zone (De Becker 2007). In this context, it is interesting to note that NGC\,6383 lies very close to the centre of the error box of the EGRET $\gamma$-ray source 3EG\,J1734\,$-$3232 (Torres et al.\ 2003).  Whilst this $\gamma$-ray source is usually associated with the supernova remnant G355.6\,$+$0.0, a contribution from NGC\,6383 and HD\,159176 cannot be ruled out (Torres et al.\ 2003).

The other remarkable early-type star in NGC\,6383 is the W\,UMa-type eclipsing binary V\,701\,Sco (= HDE\,317844, $m_V = 8.05$ at max.\ light), the second brightest member of the cluster. This system very likely consists of two almost identical and essentially unevolved B1-1.5 stars (Andersen, Nordstrom, \& Wilson 1980, Bell \& Malcolm 1987) revolving around each other in $0.762$\,days.

\section{General Cluster Properties}
A number of photometric and spectroscopic studies of NGC\,6383 can be found in the literature. Photoelectric \ubv measurements of 27 stars were presented by Eggen \cite{Eggen} whilst Th\'e \cite{The} obtained photographic photometry of 97 objects down to $V = 13.8$\,mag as well as objective prism plates. Antalov\'a \cite{Anta} reported photographic \ubv measurements of 32 stars in the neighbourhood of NGC\,6383. FitzGerald et al.\ \cite{FG} analysed photoelectric \ubv measurements of 25 stars as well as low and medium resolution spectra of 11 objects within 2\,arcmin from the cluster centre. Lloyd Evans \cite{LE1} presented photoelectric and photographic \ubv measurements of several hundred objects within 12\,arcmin from HD\,159176, along with spectral classifications for 17 stars and radial velocities for 14 of them. Th\'e et al.\ \cite{The2} discussed photoelectric observations of a sample of stars in the Walraven $VBLUW$, Cousins $VRI$ and near-infrared $JHKLM$ bands. Finally, Paunzen, Netopil \& Zwintz \cite{Paunzen} presented Str\"omgren {\it uvby} CCD photometry of 272 stars. Most authors adopted the numbering scheme introduced by Eggen \cite{Eggen} for the cluster core and extended to the outer cluster regions by Th\'e \cite{The} and Lloyd Evans \cite{LE1}\footnote{In SIMBAD, these identifiers are noted NGC\,6383\,NNN where NNN is the 3 digit number of the star.}. A somewhat different numbering scheme for the cluster core was introduced by FitzGerald et al.\ \cite{FG}\footnote{In SIMBAD, these identifiers are noted Cl$*$\,NGC\,6383\,FJL\,NN where NN is the 2 digit number of the star.}. Hereafter we use the notations `Th\'e \#' for the former and `FJL \#' for the latter. Paunzen et al.\ (2007) adopted the numbering system from the WEBDA database ({\tt http://www.univie.ac.at/webda}). In Sect.\,\ref{pms}, we also use the RDGPS numbering convention introduced by Rauw et al.\ \cite{Rauw} to designate X-ray sources in the NGC\,6383 cluster.

\subsection{The Reddening and the Cluster Distance}
Most investigations based on photometric and spectroscopic studies
agree within the errors about the reddening: Eggen \cite{Eggen}, FitzGerald et al.\ \cite{FG}, Lloyd Evans \cite{LE1}, Th\'e et al.\ \cite{The2} and Paunzen et al.\ \cite{Paunzen} inferred an $E(\bv)$ colour excess of 0.30, $0.33 \pm 0.02$, $0.35$, $0.30 \pm 0.01$ and $0.29 \pm 0.05$ respectively. Little or no spatial variation of the reddening was found over the central part of the cluster. The only exception is the study of Antalov\'a \cite{Anta} who suggested a scatter in $A_V$ between 0.6 and 1.3.

Lloyd Evans \cite{LE1} noted that the cluster is probably located right in front of a dust absorption cloud. In fact, some redder stars in the field were found to have a distance modulus only slightly larger than the one of the cluster. There exists also a region of apparently higher stellar density where the dust cloud might be thinner, thus allowing us to see some background objects. In the atlas of dark clouds of Dobashi et al.\ \cite{Dobashi}, NGC\,6383 appears indeed located on a thin bridge connecting the two giant cloud complexes TGU\,2164 and TGU\,2190 (see Fig.\,18-1-9 of Dobashi et al.\ 2005). This position corresponds to the intersection of two large cavities in the Galactic plane which might have been created by the action of early-type stars in NGC\,6383 and other young clusters in this region.

While Trumpler \cite{Trumpler} estimated a distance of 2.1\,kpc for NGC\,6383, more recent studies usually agree about a smaller distance in the range between $1.2$ and $1.5$\,kpc. Distance determinations based on colour-magnitude diagrams are provided by Eggen (1961, $d = 1.25$\,kpc), FitzGerald et al.\ (1978, $d = 1.5 \pm 0.2$\,kpc), Lloyd Evans (1978, $d = 1.32$\,kpc), Th\'e et al.\ (1985, $d = 1.4 \pm 0.15$\,kpc) and Paunzen et al.\ (2007, $d = 1.7 \pm 0.3$\,kpc). Using photoelectric measurements of the H$\beta$ index of six likely cluster members with $m_V$ in the range 5.7 -- 11.3, Graham \cite{Graham} inferred a distance modulus of $10.68 \pm 0.54$ ($d = 1.37 \pm 0.34$\,kpc). Accounting for the binarity of HD\,159176 and adopting $R_V = 3.1$ (instead of 3.0 as used by Graham 1967), one obtains a distance modulus of $10.78 \pm 0.34$ ($d = 1.43 \pm 0.22$\,kpc). A nice confirmation of these distance estimates comes from the spectroscopic and photometric analysis of the eclipsing binary system V\,701\,Sco by Bell \& Malcolm \cite{BM} who derived a distance of $1240 \pm 190$\,pc. Finally, Rastorguev et al.\ \cite{Rastorguev} used a statistical parallax technique based on the kinematical parameters (proper motions and radial velocities) to derive a cluster distance of 1.18\,kpc. To our knowledge, the lowest distance estimate is the one by Kharchenko et al.\ \cite{Kharchenko} who quote 985\,pc, although it is not fully clear what technique was used to arrive at this number. In summary, the distance to NGC\,6383 thus appears rather well determined at $1.3 \pm 0.1$\,kpc.

\subsection{Proper Motion and Cluster Membership}
The core of NGC\,6383 is rather compact (see Fig.\ \ref{optical}). Kharchenko et al.\ \cite{Kharchenko} derived a core radius (defined as the radius where the decrease of stellar surface density drops abruptly) of 4.8\,arcmin and a corona radius (where the surface density becomes equal to the average of the surrounding field) of 15\,arcmin.

The cluster membership of individual stars has been discussed using photometric criteria (e.g.\ Lloyd Evans 1978), radial velocities and proper motions. For instance, stars FJL \# 3 and 21 of FitzGerald et al.\ \cite{FG} are likely foreground B9\,IV and F stars respectively.

Several authors have investigated the cluster proper motion. Using {\it Hipparcos} data, Rastorguev et al.\ \cite{Rastorguev} inferred $\mu_{\alpha}\,\cos{\delta} = 2.0$\,mas\,yr$^{-1}$ and $\mu_{\delta} = -0.6$\,mas\,yr$^{-1}$. Using the same {\it Hipparcos} measurements, Baumgardt, Dettbarn, \& Wielen \cite{Baumg} obtained $\mu_{\alpha}\,\cos{\delta} = 2.70$\,mas\,yr$^{-1}$ and $\mu_{\delta} = -0.84$\,mas\,yr$^{-1}$, while Kharchenko et al.\ \cite{Kharchenko} derived $\mu_{\alpha}\,\cos{\delta} = 1.66 \pm 0.40$\,mas\,yr$^{-1}$ and $\mu_{\delta} = -1.56 \pm 0.34$\,mas\,yr$^{-1}$. Dias, L\'epine, \& Alessi \cite{Dias} used the Tycho2 catalogue to estimate $\mu_{\alpha}\,\cos{\delta} = 1.58 \pm 1.64$\,mas\,yr$^{-1}$ and $\mu_{\delta} = -2.01 \pm 1.64$\,mas\,yr$^{-1}$.

Dias et al.\ \cite{Dias} further derived membership probabilities for 41 individual stars brighter than $V_T = 12.2$\,mag, and found that 14 of them are most probably members (P $\geq 61\%$), 6 are possible members (P $\in [14,61]\%$) and the remainder are probable field stars. However, these results critically depend on the actual value of the proper motions and we have seen above that the various determinations are only in coarse agreement.

\section{The Star Formation Activity in NGC\,6383 \label{pms}}
Eggen \cite{Eggen} drew attention to the similarity of NGC\,6383 with NGC\,2264 in the sense that the colour-magnitude diagrams of the two clusters show a rather normal main sequence down to spectral type about A0 whilst stars of later spectral types are located above the main sequence. This feature was considered as the signature of a population of pre-main sequence (PMS) stars. FitzGerald et al.\ \cite{FG} identified eight likely cluster members that fall in the region of PMS objects in the colour-colour and colour-magnitude diagram. These authors derived a cluster age of $1.7 \pm 0.4$\,Myr. Given the rather narrow locus of potential PMS stars, they argued that star formation must have been quite coeval, except for HD\,159176 for which they inferred an age of $2.8 \pm 0.5$\,Myr, suggesting that this star triggered the formation of lower mass stars in NGC\,6383\footnote{The 5.1\,Myr cluster age proposed by Kharchenko et al.\ \cite{Kharchenko} should be considered with caution since this value relies only on the massive binary HD\,159176.}. On the other hand, Lloyd Evans \cite{LE1} suggested that the formation of lower mass stars ceased prematurely after the formation of the central cluster of massive stars, resulting in a relative lack of faint stars and the absence of T\,Tauri stars as bright as those found in regions of continuous star formation. Despite these conflicting interpretations, NGC\,6383 is an extremely interesting place to study the interplay between low- and high-mass star formation and it is therefore important to improve our knowledge about the PMS population of this cluster. A list of 44 PMS candidates in NGC\,6383 selected from their location in several colour-magnitude diagrams was recently presented by Paunzen et al.\ \cite{Paunzen}.

The problem with identifying PMS stars from classical colour-magnitude diagrams is the heavy contamination by foreground field stars. Red stars with $B - V \geq 0.7$ are essentially lost in the large population of field stars. Since the original work of Eggen, various other techniques have been applied to identify pre-main sequence stars in NGC\,6383 and we review below the main results.

\begin{figure}[htb]
\centering
\includegraphics[draft=false,width=10cm]{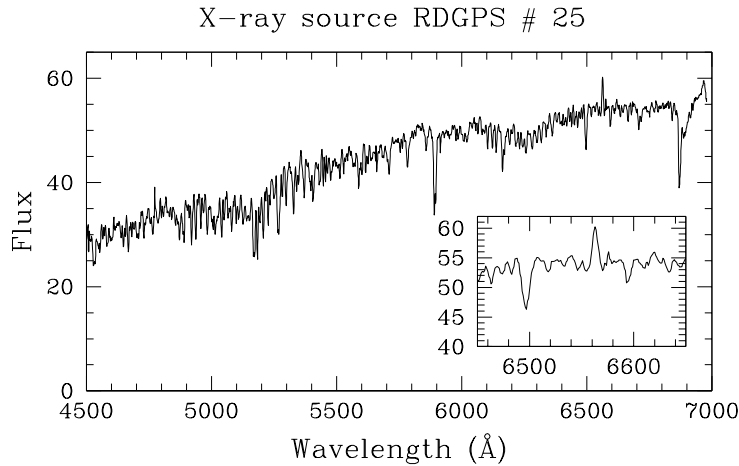}
\caption{Medium-resolution spectrum of the optical counterpart of X-ray source RDGPS \# 25 (Rauw et al.\ 2003) that we classify as K0\,V. The insert shows a zoom on the weak H$\alpha$ emission (from Rauw et al.\ 2008).\label{h-alpha}}
\end{figure}
\subsection{Emission Line Objects}
Using objective prism spectra, Th\'e \cite{The} derived approximate spectral types for 46 objects, but failed to find stars with strong H$\alpha$ emission down to $V = 16.0$\,mag. FitzGerald et al.\ \cite{FG} noted H$\beta$ emission from star FJL \# 24 on one out of six observations. They accordingly classified this star as B8\,Vne and suggested that it was a flare star undergoing the final stages of PMS contraction.
Lloyd Evans \cite{LE1} pointed out that HDE\,317861\footnote{Lloyd Evans \cite{LE1} identified this star as Th\'e \# 75, while it should actually be Th\'e \# 76.} (B3-5\,Vne) displayed a pronounced H$\beta$ emission in their observation.

Th\'e et al.\ \cite{The2} found that star FJL \# 4 (= V\,486\,Sco, A5\,IIIp) displays H$\alpha$, He\,{\sc ii} $\lambda$\,4686 and $[$O\,{\sc iii}$]$ emissions. This star is a probable Herbig Ae star. The H$\alpha$ emission of FJL \#4 was also confirmed by van den Ancker \cite{vdA} who classified it as A5\,IIIe.

Our knowledge about line emission in cooler stars is much more limited. We recently observed a sample of PMS candidates selected from their X-ray emission (Rauw et al.\ 2003). The counterparts of X-ray sources RDGPS \# 20, 25 and 57 classified as K4-5\,V-III, K0\,V and K4\,III respectively were found to display H$\alpha$ in slight emission (Rauw et al.\ 2008, in preparation and Fig.\,\ref{h-alpha}). In addition, we are in the process of collecting and analysing deep broad-band and H$\alpha$ CCD images to identify fainter H$\alpha$ emission stars that escaped detection during previous surveys (Rauw et al.\ 2008).

\subsection{Infrared Excess Emission}
Th\'e et al.\ \cite{The2} and van den Ancker et al.\ \cite{vdA} identified three stars (FJL \#4, 5 and 6) that fall above the ZAMS in the colour-magnitude diagram and show a considerable near-IR excess. While FJL \# 4 is the A5\,IIIp emission line star discussed above, stars \#5 and 6 were classified as A2\,Vep and A respectively. According to Th\'e et al.\ \cite{The2}, the near-IR excess of FJL \#4 and 5 is due to thermal emission by circumstellar dust grains that absorb the UV and visible light of the central star. From their analysis, they derived dust shell temperatures of 1700 and 1350\,K respectively for FJL \# 4 and 5.

A few more stars, FJL \# 18 (G0\,V), 20 (B8\,V) and 24 (see above), display a modest near-IR excess (van den Ancker et al.\ 2000).

\begin{figure}[htb]
\begin{center}
\includegraphics[width=\textwidth,draft=False]{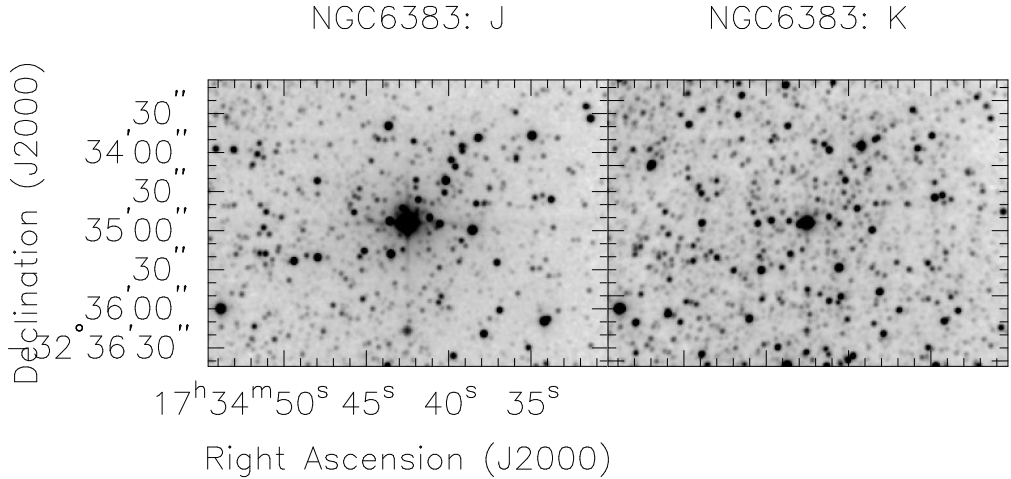}
\end{center}
\caption{Near-IR images of the core of NGC\,6383 as obtained from the 2MASS survey. Note the huge number of sources that appear strongest in the $K_s$ filter.\label{2mass}}
\end{figure}

The 2MASS images of the region near NGC\,6383 reveal a large number of sources that appear most obviously in the $K_s$ filter (see Fig.\,\ref{2mass}). Some of them are rather heavily absorbed objects without an optical counterpart, but associated to an X-ray source (see below) and could be either background objects or deeply embedded PMS objects.

\subsection{Variable Pre-Main Sequence Stars}
PMS stars often display a high degree of spectroscopic and photometric variability over a wide range of timescales from minutes to months. Several variables have been found from snapshot observations of NGC\,6383. FitzGerald et al.\ \cite{FG} noted that star FJL \#10 ($m_V = 15.3$) displayed photometric variability of 0.3\,mag. Lloyd Evans \cite{LE1} identified six stars that display photometric variability with one object (his star V1) showing a variation of more than 2\,mag. Lloyd Evans \cite{LE1} suggested that these objects are RW\,Aur or T\,Tauri variables. Finally, Th\'e et al.\ \cite{The2} drew attention to the star FJL \#3 that apparently displays a changing spectral type.

A systematic search for PMS variables was performed by Zwintz et al.\ \cite{zwintz1,zwintz2}. These authors obtained time series of $B$ and $V$ photometry of NGC\,6383 over two weeks. Out of 15 cluster members that are located in the classical instability strip, two A -- F type PMS stars were found to display $\delta$ Scuti-like pulsations. Interestingly enough, one of them was again the A5\,IIIp Herbig Ae/Be star FJL \# 4 for which five different periods between 1.24 and 2.89\,hours could be identified as radial and non-radial $p$-mode pulsations (see Zwintz, Guenther \& Weiss 2007). For Th\'e \# 55, a single frequency corresponding to 1.26\,hours was detected by Zwintz et al.\ \cite{zwintz1,zwintz2}. A third object (Th\'e \# 54) was classified as a suspected PMS pulsator. Several other stars were found to display photometric variations. Some of them are probably foreground or background objects unrelated to the cluster, but four objects were classified as T\,Tauri stars or T\,Tauri candidates by Zwintz et al.\ \cite{zwintz2}.

\begin{figure}[htb]
\begin{center}
\includegraphics[width=\textwidth,draft=False]{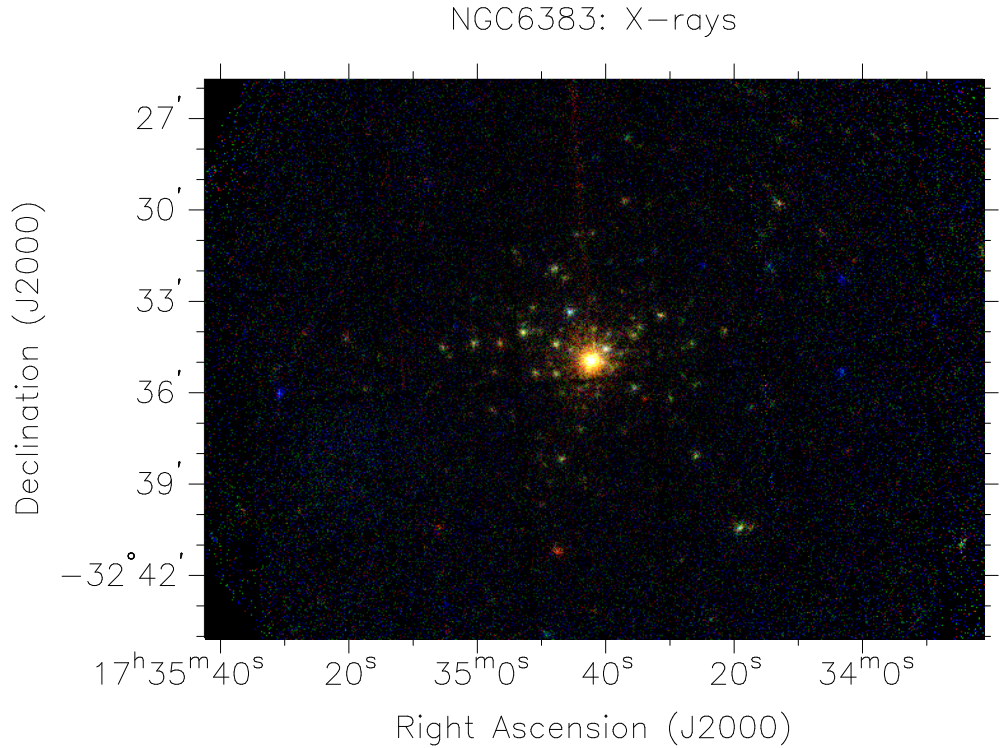}
\end{center}
\vspace*{-5mm}
\caption{Three colour X-ray image of NGC\,6383 as obtained from the {\it XMM-Newton} EPIC images of Rauw et al.\ \cite{Rauw}. The colours correspond to different energy bands: red, green and blue stand for emission in the $[0.5,1.0]$\,keV, $[1.0,2.0]$\,keV and $[2.0,8.0]$\,keV energy domains respectively. The brightest source in the image is HD\,159176.\label{xrays}}
\end{figure}

\subsection{X-ray Selected PMS Candidates}
Relatively bright, hard and variable X-ray emission is another common feature of classical (cTTs) and weak-line T\,Tauri stars (wTTs). In addition to HD\,159176, Rauw et al.\ \cite{Rauw} detected 76 secondary X-ray sources in the {\it XMM-Newton} field of view of NGC\,6383. These sources are strongly concentrated towards the cluster centre (see Fig.\ \ref{xrays}). The coordinates of the X-ray sources were cross-correlated with various optical and near-IR catalogues. Most sources have quite faint optical counterparts indicating that they must have a rather large $L_X/L_{vis}$ ratio. While three PMS candidates of FitzGerald et al.\ \cite{FG} FJL \# 11, 23 and 24 were seen in X-rays, none of the variables of Lloyd Evans \cite{LE1} was detected. The three A-type stars (FJL \# 4, 5 and 6) that present a strong near-IR excess were also not detected.

The brightest secondary X-ray sources have spectra that can be represented by a two-component thermal plasma model with plasma temperatures in the range 6 -- 12\,MK and 25 -- 60\,MK for the cooler and hotter component respectively. Source RDGPS \#50 (Rauw et al.\ 2003) displayed a flare towards the end of the observation with a significant hardening of its emission. These properties suggest that these objects are indeed low-mass PMS stars. Only two X-ray sources were found to have a counterpart with near-IR colours indicating excess emission. This rather low fraction of objects with evidence for circumstellar material suggests that most of the X-ray selected PMS candidates are in fact wTTs rather than cTTs. This is confirmed by the rather low incidence of H$\alpha$ emitters found in the spectroscopic follow-up observation of the X-ray selected PMS candidates (Rauw et al.\ 2008, in preparation and Fig.\,\ref{h-alpha}). Finally, 16 X-ray sources have possible near-IR counterparts with $A_V$ up to 28\,mag, substantially larger than the cluster reddening. These could be either deeply embedded Class I protostars or background objects unrelated to the cluster.\\

Table\,\ref{over} provides an overview of the properties of the PMS candidate stars discussed hereabove in view of the various selection criteria that have been used.
\begin{table}[!htbp]
\caption{Summary of the multiwavelength properties of the PMS candidates in NGC\,6383 discussed in this review. The numbers in the last four columns indicate the references where these features were reported. Note that there are many more X-ray selected PMS candidates, but we decided to focus on those that either display a flaring activity or exhibit a weak H$\alpha$ emission in their optical spectrum. \label{over}}
\begin{center}
\begin{tabular}{l l l c r c c c c}
\hline
\multicolumn{1}{c}{Object} & \multicolumn{1}{c}{$\alpha_{2000}$} & \multicolumn{1}{c}{$\delta_{2000}$} & \hspace*{-3mm} Spect.\ type & \multicolumn{1}{c}{$V$} & Em. & IR & Var. & X \\
\hline
HDE\,317861 & \hspace*{-3mm} 17:34:02 & \hspace*{-3mm} $-32$:40:40 & \hspace*{-3mm} B3-5\,Vne & \hspace*{-3mm} 9.9 & 1 & - & - & - \\
FJL \# 4  & \hspace*{-3mm} 17:34:38 & \hspace*{-3mm} $-32$:36:19 & \hspace*{-3mm} A5\,IIIp & \hspace*{-3mm} 13.0 & 2,3 & 2,3 & 4 & - \\
FJL \# 5  & \hspace*{-3mm} 17:34:34 & \hspace*{-3mm} $-32$:34:36 & \hspace*{-3mm} A2\,Vep  & \hspace*{-3mm} 12.9 & - & 2,3 & - & - \\
FJL \# 6  & \hspace*{-3mm} 17:34:39 & \hspace*{-3mm} $-32$:33:59 & \hspace*{-3mm} A6       & \hspace*{-3mm} 13.8 & - & 2,3 & - & - \\
FJL \# 10 & \hspace*{-3mm} 17:34:40 & \hspace*{-3mm} $-32$:33:45 &          & \hspace*{-3mm} 15.3 & - & - & 5 & - \\
FJL \# 11 & \hspace*{-3mm} 17:34:42 & \hspace*{-3mm} $-32$:33:52 &          & \hspace*{-3mm} 15.1 & - & - & 4 & 6 \\
FJL \# 18 & \hspace*{-3mm} 17:34:37 & \hspace*{-3mm} $-32$:35:24 & \hspace*{-3mm} G0\,V    & \hspace*{-3mm} 13.4 & - & 2,3 & - & - \\
FJL \# 20 & \hspace*{-3mm} 17:34:38 & \hspace*{-3mm} $-32$:33:49 & \hspace*{-3mm} B9\,IV   & \hspace*{-3mm} 11.4 & - & 2,3 & - & - \\
FJL \# 23 & \hspace*{-3mm} 17:34:48 & \hspace*{-3mm} $-32$:34:22 & \hspace*{-3mm} G5\,V$^*$& \hspace*{-3mm} 13.8 & - & - &  & 6 \\
FJL \# 24 & \hspace*{-3mm} 17:34:48 & \hspace*{-3mm} $-32$:35:20 & \hspace*{-3mm} B8\,Vne  & \hspace*{-3mm} 11.4 & 5 & 2,3 & - & 6 \\
Th\'e \# 55 & \hspace*{-3mm} 17:34:49 & \hspace*{-3mm} $-32$:37:21 &        & \hspace*{-3mm} 12.9 & - & - & 4 & - \\
LE V1 & \hspace*{-3mm} 17:34:06 & \hspace*{-3mm} $-32$:32:53 &  &  & - & - & 1 & - \\
RDGPS \# 20 & \hspace*{-3mm} 17:34:26 & \hspace*{-3mm} $-32$:37:59 & \hspace*{-3mm} K4-5\,V-III & \hspace*{-3mm} 15.4 & 7 & - & - & 6 \\
RDGPS \# 25 & \hspace*{-3mm} 17:34:32 & \hspace*{-3mm} $-32$:33:26 & \hspace*{-3mm} K0\,V & \hspace*{-3mm} 14.5 & 7 & - & - & 6 \\
RDGPS \# 50 & \hspace*{-3mm} 17:34:46 & \hspace*{-3mm} $-32$:33:18 & \hspace*{-3mm} $^{**}$&      & - & - & - & 6 \\
RDGPS \# 57 & \hspace*{-3mm} 17:34:48 & \hspace*{-3mm} $-32$:31:55 & \hspace*{-3mm} K4\,III & \hspace*{-3mm} 15.2 & 7 & - & - & 6 \\
\noalign{\smallskip}
\tableline
\noalign{\smallskip}
\multicolumn{9}{l}{\parbox{0.9\textwidth}{\footnotesize
(1) = Lloyd Evans \cite{LE1}, (2) = Th\'e et al.\ \cite{The2}, (3) = van den Ancker et al.\ \cite{vdA}, (4) Zwintz et al.\ \cite{zwintz2}, (5) = FitzGerald et al.\ \cite{FG}, (6) = Rauw et al.\ \cite{Rauw}, (7) = Rauw et al.\ (2008, in prep.)}}\\[1ex]
\multicolumn{9}{l}{\parbox{0.9\textwidth}{\footnotesize
$^*$ possible foreground object.}}\\[1ex]
\multicolumn{9}{l}{\parbox{0.9\textwidth}{\footnotesize
    $^{**}$ there are two optical counterparts inside the positional
    error box of the X-ray source. }}
\end{tabular}
\end{center}
\vspace*{-5mm}
\end{table}

\begin{figure}[htbp]
\centering
\includegraphics[draft=False,width=0.9\textwidth,angle=0]{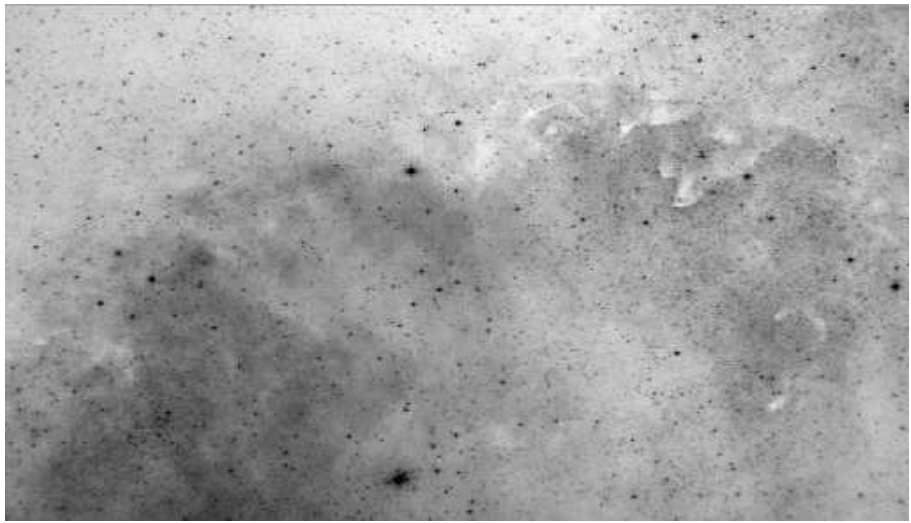}
\caption{ESO red Schmidt film of the region north of NGC\,6383. The cluster is in the lower middle of the image with HD\,159176 appearing as the brightest source. The size of the image is $2.0^{\circ} \times 1.2^{\circ}$.\label{nextgen}}
\end{figure}

\subsection{Second-generation Star Formation?}
Through the combined action of their stellar winds and their intense radiation field, early-type stars can sweep up and compress the material of an ambient or nearby molecular cloud. This interaction is thought to trigger the formation of new generations of stars and evidence for such a process has been found in the periphery of several young open clusters (e.g.\ Westerlund\,2, Whitney et al.\ 2004; NGC\,6604, Reipurth, this volume).

ESO red Schmidt films reveal an extended shell like structure to the north of NGC\,6383 (see Fig.\,\ref{nextgen}). This shell features several dense clouds and cometary globules. The high-opacity structure in the upper middle probably corresponds to the dark nebula LDN\,1734 catalogued by Lynds \cite{Lynds}. These structures are very much reminiscent of the dust pillars seen in well known examples of second-generation star formation sites (e.g.\ Walborn 2002) and could hence be the result of the interaction between the winds and radiation of the massive stars and their primordial cloud. To our knowledge, none of the structures around NGC\,6383 has been investigated in details but they are ideal targets for future studies to search for young stars.\\

{\bf Acknowledgements.}
It is a pleasure to thank Dr.\ Bo Reipurth for inviting us to prepare this chapter, for extremely valuable suggestions and for providing us with Fig.\, 5.
We are grateful to Dr.\ Ya\"el Naz\'e for assistance in the preparation of Figs.\  1, 3 and 4 and to the referee Dr.\ Konstanze Zwintz for carefully reading our manuscript. We acknowledge financial support from the FRS/FNRS (Belgium), as well as through the XMM and INTEGRAL PRODEX contract and contract P5/36 PAI (Belspo).

\end{document}